\documentclass[aip,pop,onecolumn,preprintnumbers,amsmath,amssymb,english]{revtex4-1}

\usepackage{graphicx}
\usepackage[nooneline]{subfigure} 
\usepackage{dcolumn}
\usepackage{bm}
\usepackage{color}
\usepackage{textcomp}
\usepackage{babel}
\usepackage{csquotes}
\MakeOuterQuote{"}

\usepackage{comment}

\begin{document}

\pagestyle{empty} 

\title{On The Origin of Super-Hot Electrons from Intense Laser Interactions with Solid Targets having Moderate Scale Length Preformed Plasmas}

\author{A.G. Krygier$^{1}$, D.W. Schumacher$^{1}$, R.R. Freeman$^{1}$}
\affiliation{$^{1}$Physics Department, The Ohio State University, Columbus, Ohio, 43210, USA}
\date{\today}

\begin{abstract}
We use particle-in-cell modeling to identify the acceleration mechanism responsible for the observed generation of super-hot electrons in ultra-intense laser-plasma interactions with solid targets with pre-formed plasma.  We identify several features of direct laser acceleration (DLA) that drive the generation of super-hot electrons. We find that, in this regime, electrons that become super-hot are primarily injected by a looping mechanism that we call loop-injected direct acceleration (LIDA). 
\end{abstract}

\maketitle

\section{Introduction}

Over the last decade, laser facilities with peak intensity $I>10^{20}\frac{W}{cm^{2}}$ have enabled the study of a variety of exciting applications including ion acceleration \textcolor{blue}{\cite{Clark:2000,Maksimchuk:PRL2000,Snavely:PRL2000}}, x-ray generation \textcolor{blue}{\cite{Murnane:1989,Kneip:ProcSPIE2009}}, and laboratory astrophysics \textcolor{blue}{\cite{Remington:1999,Chen:PRL2009}}. These applications are driven by the relativistic electron beams generated by ultraintense short pulse lasers interacting with plasma and, in general, are enhanced by maximizing the hot electron current and energies. It is therefore important to understand the precise nature of how the electron beams are generated.

There are a variety of mechanisms by which a laser can couple its energy into relativistic electrons in plasma. First is the well-known laser wake-field scheme \textcolor{blue}{\cite{Mangles:Nat2006}} where a short pulse laser interacts with low density plasma to produce a directional and high energy electron beam; however, the typical currents produced by the wake-field mechanism are lower than is generally desired for the above applications. On the other hand, solid-target interactions produce high current electron beams but suffer from broad energy \textcolor{blue}{\cite{Wilks:1992, Beg:1997, Yabuuchi:2007, HChen:POP2009, Tanimoto:2009, Link:2011}} and angular spread \textcolor{blue}{\cite{Stephens:PRE2004,Akli:PRE2012}}. In this case, the relativistic or "hot" electron population is typically generated when an intense laser interacts with the pre-formed plasma generated by amplified spontaneous emission incident on the solid target starting a few ns before the short-pulse laser. This electron population can often be characterized by a so-called "slope temperature" approximately given by the "ponderomotive energy" $kT_{p}$ 

\begin{equation} 
\begin{aligned}
kT_{p} = m_{e}c^{2}\bigg(\sqrt{1+\frac{I\lambda^{2}[\frac{W\mu m^{2}}{cm^{2}}]}{1.37*10^{18}}}-1\bigg)
\end{aligned}
\end{equation}

\noindent where $m_{e}$ is the electron mass, $c$ is the speed of light, and $\lambda$ is the laser wavelength. This scaling has been the subject of experimental \textcolor{blue}{\cite{Beg:1997,Haines:2009}}, computational \textcolor{blue}{\cite{Wilks:1992}}, and theoretical \textcolor{blue}{\cite{KlugeEnergy:2011}} efforts. This work seeks to answer the question of how the highest energies of the distribution, which we term "super-hot", are produced when an intense laser interacts with a solid target with pre-formed plasma.

The super-hot region has been the subject of both theoretical scrutiny and computer simulation. Particle-in-cell (PIC) modeling has been used to study electron acceleration by an ultraintense short pulse laser in a plasma channel \textcolor{blue}{\cite{Pukhov:1998,Pukhov:1999,Gahn:1999,Mangles:2005}}; in particular, the betatron resonance mechanism was identified \textcolor{blue}{\cite{Pukhov:1999}} as a way to enhance electron energies in a long plasma channel. Others have investigated the role of stochastic fields on electron acceleration \textcolor{blue}{\cite{MeyerterVehn:1999,Tanimoto:2003,Paradkar:2011}}. PIC modeling was used to identify direct laser acceleration (DLA) as the mechanism responsible for the experimentally observed enhancement to electron and ion energy spectra when a high-contrast short-pulse laser interacts with the wall of a flat-top cone \textcolor{blue}{\cite{Gaillard:2011,Kluge:2012}}. For this case, electrons were injected directly into the intense laser field by the side of the cone wall, thus producing the enhanced spectra. PIC modeling and theory have also been used to identify an enhancement to DLA caused by the inclusion of static transverse \textcolor{blue}{\cite{Arefiev:2012}} and longitudinal \textcolor{blue}{\cite{Robinson:PRL2013}} electric fields. Compared to the free electron case, the static electric fields, like those in a laser-formed channel, can help slow the dephasing rate of the electron from the laser and lead to enhanced energies. 

In this work, we use PIC modeling to investigate one of the most common and fundamental scenarios: a short-pulse laser incident on a flat, solid target with moderate scale-length pre-formed plasma. We identify all of the super-hot electrons (a small fraction of the total population) at the end of the simulation and reconstruct the histories of a large sample of the super-hots. In doing this we have found that a large majority of the super-hots are produced by a mechanism we call "loop-injected direct acceleration" (LIDA) which injects electrons into the peak intensity of the laser enabling large energy gain via DLA.

\begin{figure}
\includegraphics[width=0.45\textwidth,natwidth=1056,natheight=364]{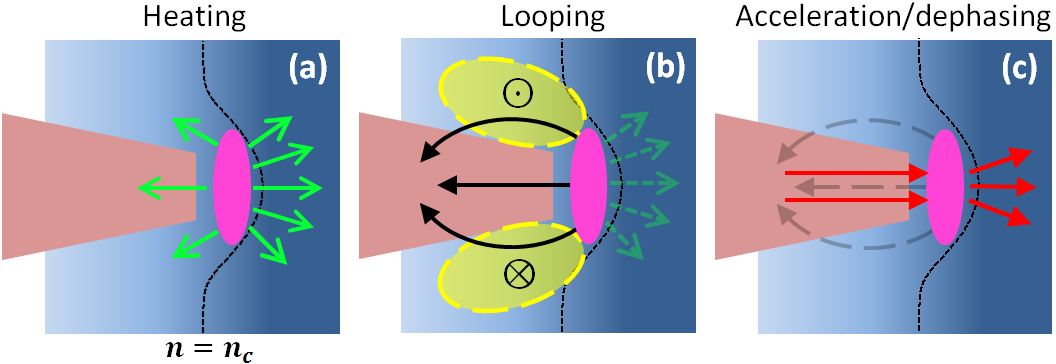}
\caption{(color online). Sketch depicting LIDA. In (a), the incoming laser heats the region (pink) near the critical surface (dashed black line) expelling electrons (green arrows). The electrons that will become super-hots follow looping trajectories (black arrows) shown in (b), shaped by large quasi-static fields (the magnetic field is indicated by yellow). The electrons complete the loop in (c) where they undergo DLA until they reach the critical surface, now with large energies, and escape into the target.}
\label{fig:loop_cartoon}
\end{figure}

The LIDA mechanism is initiated by the early part of the laser pulse heating the region near the critical surface whose general area is indicated in pink in FIG. \textcolor{blue}{\ref{fig:loop_cartoon}}. The angular distribution of the electrons in this heated region is broad and electrons leave and enter the region in all directions; the green arrows in FIG. \textcolor{blue}{\ref{fig:loop_cartoon}.a} indicate the leaving electrons. We find that the electrons that will become super-hots, however, leave this region following the looping paths indicated by the black arrows in FIG. \textcolor{blue}{\ref{fig:loop_cartoon}.b}. These paths are shaped by large quasi-static fields (the orientation and approximate extent of the magnetic field is shown in yellow in FIG. \textcolor{blue}{\ref{fig:loop_cartoon}.b}). The loop is completed when the super-hots are injected into the laser and start undergoing DLA as shown in FIG. \textcolor{blue}{\ref{fig:loop_cartoon}.c}. The electrons are then accelerated until they reach the critical surface where they decouple from the laser and escape with high energy into the target. \emph{Our studies indicate that LIDA dominates the acceleration of electrons with final energy at a large multiple of the ponderomotive energy.}

\section{Free Electron Interaction With a Laser}

Before discussing our simulation results, we consider the motion of a free electron initially at rest which interacts with a linearly polarized electromagnetic plane wave whose temporal and spatial features are an arbitrary function of $\phi$. The vector potential $A$ is given by

\begin{equation} 
\begin{aligned}
\vec{A} & =A(\phi)\hat{x}\\ 
\phi & =\omega t-kz 
\end{aligned}
\end{equation}

\noindent where $\omega$ is the laser frequency, $t$ is the time, $k=2\pi/\lambda$ is the wavenumber, and $z$ is the position along the propagation direction.  The vector potential is related to the electric field $E$ and the magnetic field $B$ through 

\begin{equation}
\begin{aligned}
\vec{E} & =-\frac{\partial\vec{A}}{\partial t}=-\omega\frac{\partial A}{\partial\phi}\hat{x}\\
\vec{B} & =\vec{\nabla}\times\vec{A}=-k\frac{\partial A}{\partial\phi}\hat{y}
\end{aligned}
\end{equation}

\noindent Using these relations the Lorentz force law  

\begin{equation}
\frac{d\vec{p}}{dt}=-e\left[\vec{E}+\vec{v}\times\vec{B}\right]
\end{equation}

\noindent where $v$ is the electron velocity, can be easily integrated. Substituting in the normalized vector potential $a = eA/m_{e}c$ where $e$ is the elementary charge, we get the well-known solutions for the electron's momentum

\begin{equation}
\begin{aligned}
p_{x} & =(a-a_{i})m_{e}c\\
p_{z} & =(a-a_{i})^{2}\frac{m_{e}c}{2}
\end{aligned}
\label{eq:momenta_rel}
\end{equation}

\noindent where $p_{x}$ and $p_{z}$ are the transverse and longitudinal momentum of the electron and the subscript $i$ indicates the initial condition. Therefore, the momentum of an electron evolving under theses conditions is fully defined by the difference between the initial and the current values of the vector potential. When $a_{i}=0$ the difference $a-a_{i}$ is limited to $\pm a_{0}$ but when $a_{i}=\pm a_{0}$ the difference can be as much as $\pm 2a_{0}$; accordingly, the latter case enables much higher energies. The normalized energy rate of the electron is given by

\begin{equation}
\begin{aligned}
\frac{d\gamma}{dt} &= -\frac{e}{m_{e}c^{2}}\vec{v}\cdot\vec{E} \\
&= -\frac{e}{m_{e}c^{2}}v_{x}E_{x}
\end{aligned}
\end{equation}

\noindent where $\gamma = 1 / \sqrt{1-(v/c)^{2}}$ is the relativistic factor. This relation can be easily integrated and, in combination with the Lorentz force law, gives the relation for the kinetic energy

\begin{equation}
\begin{aligned}
\gamma &= \frac{p_{z}}{m_{e}c} + 1. \\
U_{e} &= (a-a_{i})^{2}\frac{m_{e}c^{2}}{2}
\end{aligned}
\end{equation}

\noindent where $U_{e}$ is the electron kinetic energy which scales with the quantity $(a-a_{i})^{2}$. 

This simple analytical solution gives rise to two distinct cases: first, the non-injection case where $a_{i}=0$ and second, the injection case where $|a_{i}|>0$ (recall that $a$ and $a_{i}$ are signed, unlike $a_{0}$). In the first case, the range of the electron's energy is limited to $U_{e}= [0,a_{0}^{2}m_{e}c^{2}/2]$. For the conditions in our simulation ($a_{0}\simeq 21$), this gives a maximum possible energy of $U_{e,max}\simeq 112 MeV$. However, experiments and computer PIC modeling typically produce electrons with energies that are significantly higher than the non-injection limit (which of course varies with the corresponding $a_{0}$) \textcolor{blue}{\cite{Chen:PRL2009,Pukhov:1998,Kemp:PRL2012,Link:2011}}. Indeed, our PIC modeling also predicts electrons with greater energy than this limit ($U_{e,max}\simeq 200 MeV$, see FIG. \ref{fig:espec_source}). On the other hand, if there is an injection mechanism, the analytical model results in an increase in the range of the electron kinetic energy to $U_{e}= [0,2a_{0}^{2}m_{e}c^{2}]$ which corresponds to $U_{e,max} \simeq 448MeV$ for the simulation conditions. As we describe below, we find two injection mechanisms which produce the highest energy electrons in our PIC modeling: LIDA, where particles are injected into the laser by following a looping path as well as laser ionization-injection where an electron is ionized by the laser field and is immediately directly accelerated by the laser to high energy.

In practice, achieving the intense fields considered in this work is only possible by tightly focusing the short-pulse laser. The need to satisfy $\vec{\nabla}\cdot\vec{E}=0 = \vec{\nabla}\cdot\vec{B}$ gives rise to longitudinal electric and magnetic fields (which scale with $1/kw_{0}$, where $w_{0}$ is the beam waist), greatly complicating the theoretical expressions. The dynamics of an electron interacting with a focusing laser pulse have been considered analytically elsewhere \textcolor{blue}{\cite{Mcdonald:PRLcomment1998, Mora:PRLcomment1998, QuesnelMora:PRE1998, Maltsev:PRL2003, Yang:PPCF2009}}. In particular, the authors of reference \textcolor{blue}{\cite{Yang:PPCF2009}} derive detailed expressions for the momenta and position; the authors of \textcolor{blue}{\cite{Maltsev:PRL2003}} make estimates for energy gain and dephasing time by making some simplifying assumptions. Our PIC modeling includes all of the effects of the putative laser focus as well as the self-consistent fields generated by the displacement of charge in the plasma.

\section{Modeling Parameters}

\begin{figure}
\includegraphics[width=7cm,natwidth=1738,natheight=1529]{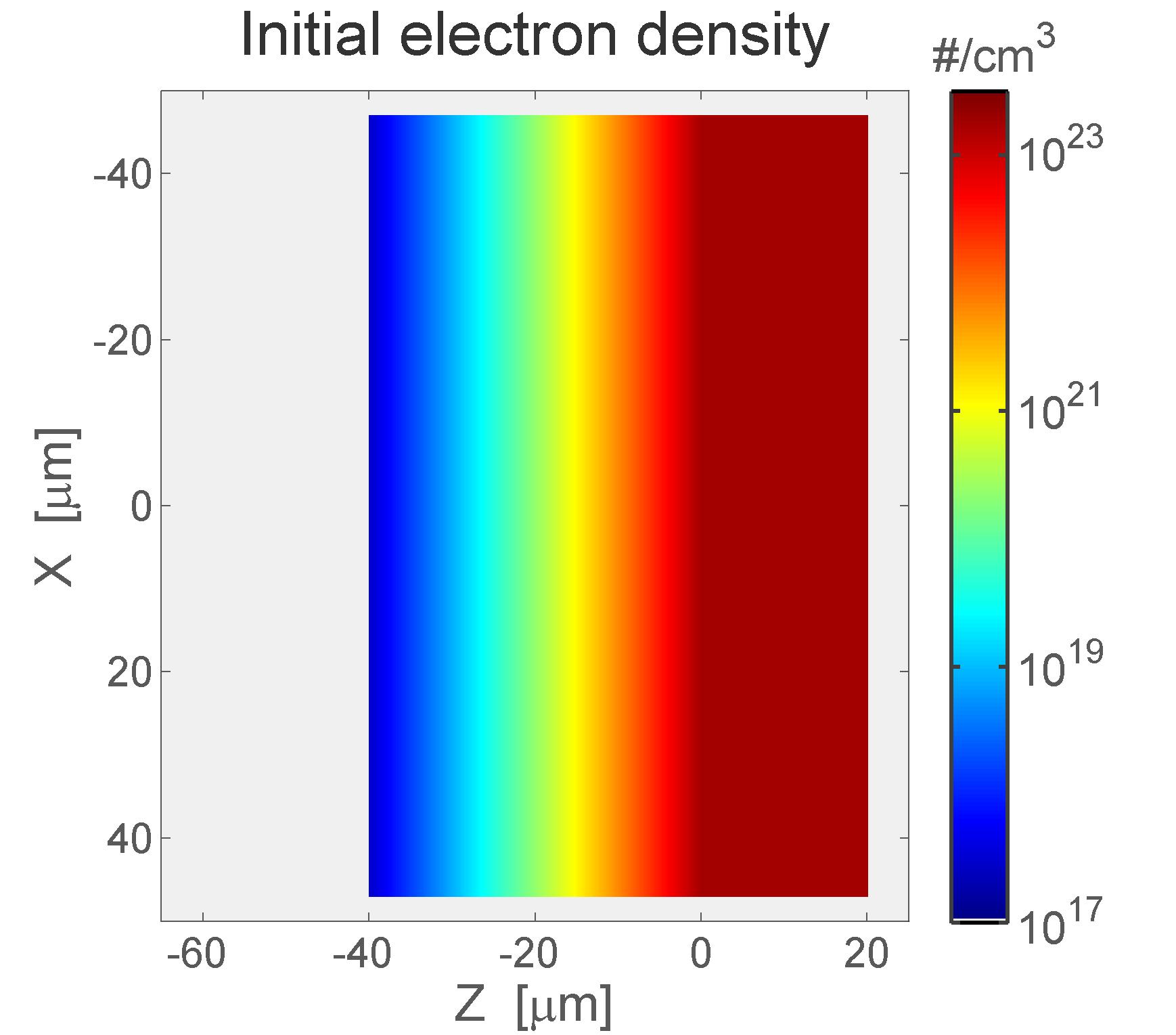}
\caption{(color online). A colorplot showing the initial plasma electron density (log scale). The $175fs$ intensity FWHM Gaussian laser enters from the $z = -65\mu m$ boundary and is focused with a vacuum intensity FWHM of $\sim 9\mu m$ at ($z=0,x=0$). The three edges of the plasma through which the laser does not pass absorb fields and particles to prevent refluxing.}
\label{fig:initial_density}
\end{figure}

We model a $\lambda=1\mu m$ Gaussian ($175fs$ and $\sim9\mu m$ temporal and spatial intensity FWHM, respectively) laser pulse with peak intensity $I=6*10^{20}\frac{W}{cm^{2}}$ (corresponding to $100J$ of 3D-equivalent laser energy) with the PIC code \textsc{lsp} \textcolor{blue}{\cite{Welch:2006}} which uses the direct implicit method with an energy conserving particle push. The transverse ($x$) polarized laser pulse propagates in the longitudinal ($z$) direction through a 2D Cartesian grid with spatial resolution $dx=dz=\frac{\lambda}{32}$ and $\sim95$ time steps per optical cycle. Entering from the left in FIG. \textcolor{blue}{\ref{fig:initial_density}}, the normally incident laser pulse propagates through $15\mu m$ of vacuum before entering a pre-formed plasma region with a $3\mu m$ (exponential) scale length and $40\mu m$ total length; a solid density region abuts the pre-plasma and extends $20\mu m$ in the longitudinal direction. The pre-plasma and solid regions extend $84\mu m$ in the transverse direction and are bounded by conducting surfaces on the three non-laser boundaries. The plasma conditions are chosen to be similar to standard experimental conditions and radiation hydrodynamics modeling results \textcolor{blue}{\cite{LePape:09,Ping:PRL2008}} while limiting the size of the simulation. As we note below, we have investigated but don't present other conditions. The Al plasma is initialized to a $5eV$ temperature and is comprised of $49$ electron and $49$ ion macro-particles per cell; the ions are initialized at $Z=+3$ and are further dynamically ionized using ADK rates \textcolor{blue}{\cite{Ammosov:1986}}, but not collisions. The importance of LIDA is robust against variation in numerical parameters such as particle count or resolution.  

In our simulations, each macro-particle is uniquely tagged allowing us to construct individual particle history tracks. First, all of the electrons with energy above $5MeV$ which cross into the solid density region and have thus escaped the laser-plasma interaction are identified. Having isolated the particles-of-interest from the much larger total population, individual position and momentum histories are reconstructed for all times in the simulation. This retrospective particle tracking method allows us to follow every particle in the simulation at all times but eventually focus only on those that become super-hots.

A sampling of 500 particles is investigated in each $5MeV$ wide bin in the range $U_{e} = 5-150MeV$; above $150MeV$ the total number of tracks in each bin is less than 500, so every track was investigated. The injection mechanism for each track is determined by inspection  and comparison with the features described below. Based on this, we are able to separate the hot electrons into two distinct injection mechanisms: those injected by LIDA and those injected by laser ionization. A third group which is made up of relatively low energy electrons is also identified. Generally, the third group has no identifiable injection mechanism and typically has a stochastic path through the LPI region before gaining its energy from the laser and/or quasi-electrostatic fields; this group is referred to as "other". 

\section{Injection and Acceleration}

\begin{figure*}
\includegraphics[width=0.9\textwidth,natwidth=4656,natheight=2556]{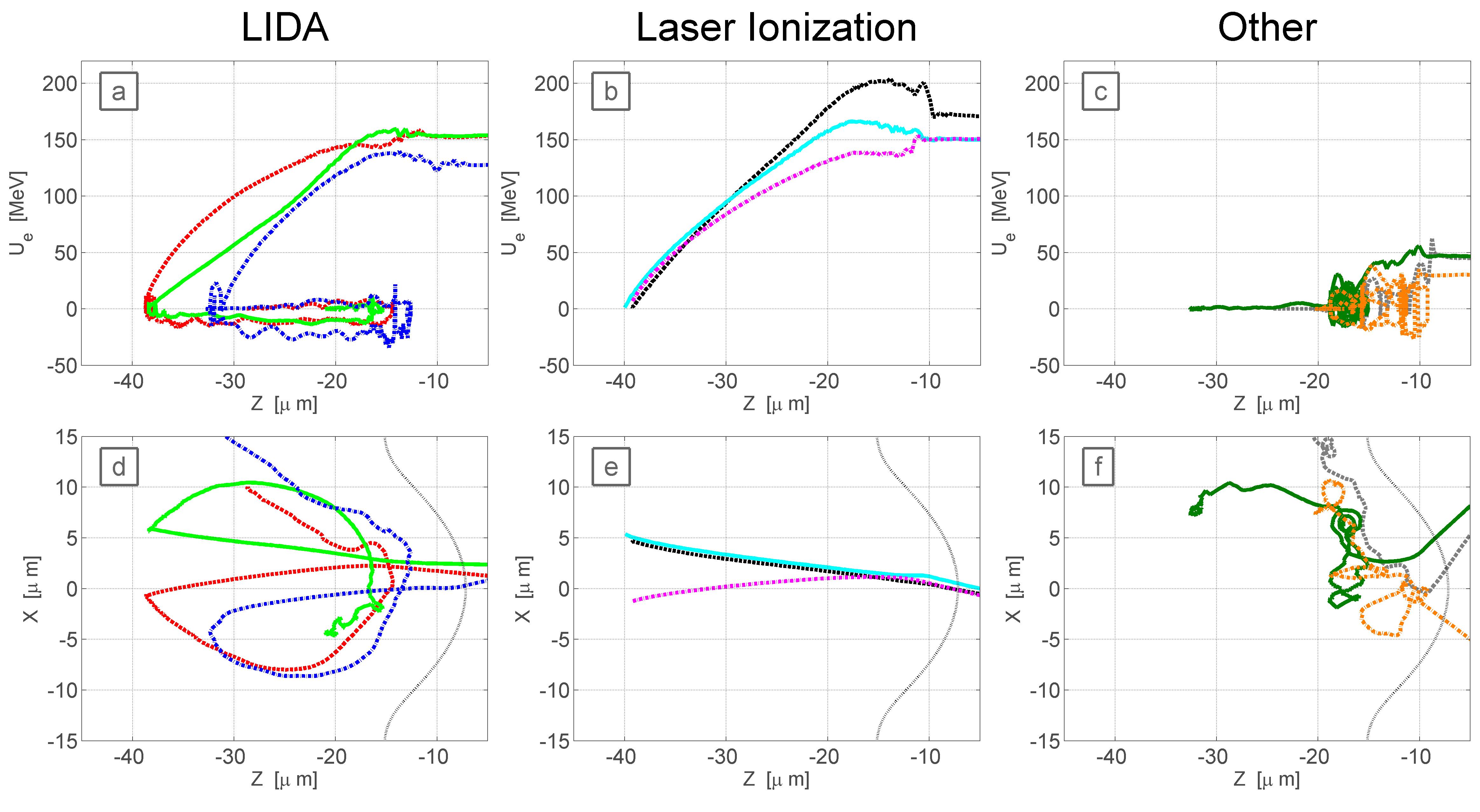}
\caption{ (color online). Nine particle tracks which highlight the characteristic features of each injection mechanism. Three LIDA tracks are plotted in red, blue, and green in \textcolor{blue}{a} and \textcolor{blue}{d}; three laser ionization-injection tracks are plotted in black, pink, and cyan in \textcolor{blue}{b} and \textcolor{blue}{e}; three other tracks are plotted in dark green, orange, and gray in \textcolor{blue}{c} and \textcolor{blue}{f}. For each case, the top row plots the kinetic energy (signed by $p_{z}$) vs. longitudinal position (\textcolor{blue}{a}, \textcolor{blue}{b}, \textcolor{blue}{c}) and the bottom row plots the trajectory (\textcolor{blue}{d}, \textcolor{blue}{e}, \textcolor{blue}{f}). In \textcolor{blue}{d}, \textcolor{blue}{e}, and \textcolor{blue}{f} the location of the critical density is shown by the dashed line. The $x$-dependent relativistic critical density is calculated using the peak field at each point in $x$ for the initial plasma conditions.}
\label{fig:sample_trajectories} 
\end{figure*}

Three individual representative particle tracks from the simulation are shown in FIG. \textcolor{blue}{\ref{fig:sample_trajectories}} for each of the two observed injection mechanisms, LIDA (\textcolor{blue}{a} and \textcolor{blue}{d}) and laser ionization-injection (\textcolor{blue}{b} and \textcolor{blue}{e}) as well as for tracks falling into the other category (\textcolor{blue}{c} and \textcolor{blue}{f}). The top three images (\textcolor{blue}{a}, \textcolor{blue}{b}, \textcolor{blue}{c}) in FIG. \textcolor{blue}{\ref{fig:sample_trajectories}} show the kinetic energy vs. longitudinal position; the bottom three images (\textcolor{blue}{d}, \textcolor{blue}{e}, \textcolor{blue}{f}) show the trajectories. The kinetic energy carries the sign of the longitudinal momentum $p_{z}$ so negative kinetic energy indicates $p_{z}<0$ (backwards moving). We now describe the two observed injection mechanisms.

\begin{figure}
\includegraphics[width=8cm,natwidth=2214,natheight=1304]{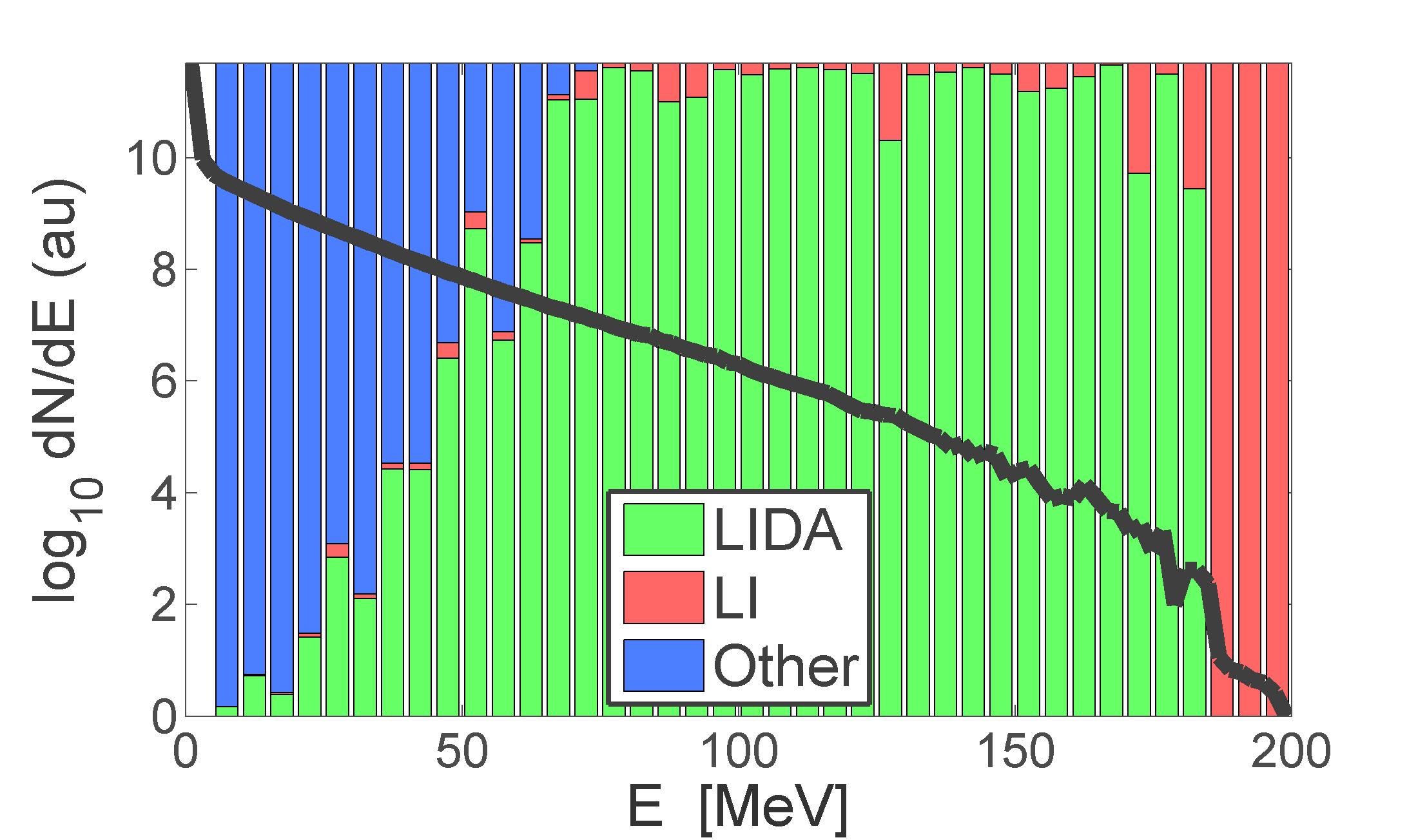}
\caption{ (color online). The modeled electron energy spectrum ($p_{z}>0$, integrated over all angles) is plotted (log scale) in black. This energy distribution is calculated using every electron crossing a plane $5\mu m$ into solid density. The colored bars form a normalized histogram (linear scale) showing the fraction of the sampled hot electrons which are injected by LIDA (green), laser ionization (LI, red), or some other irregular mechanism (blue). The histogram uses randomly sampled electrons that cross into solid density with the energy indicated. LIDA electrons account for $>41 \%$ of the total energy in the spectrum above $5MeV$.}
\label{fig:espec_source}
\end{figure}

The relative proportions of each of the three groups as a function of energy are shown by the normalized linear histogram in FIG. \textcolor{blue}{\ref{fig:espec_source}}. For the combination of intensity and pre-plasma in this simulation, all of the particles we tracked whose energy exceeds $75MeV$ required one of the two injection mechanisms. Below $75MeV$, the role of a clear injection mechanism starts to diminish in favor of other mechanisms as indicated in the figure. This finding is consistent with the above free electron theory. Altogether, LIDA and laser ionization-injection account for $>41\%$ and $1\%$ ($36\%$ and $1\%$) of the electron energy (charge), respectively, that gets coupled into electrons with energy $>5MeV$ and travels into the solid target.

\subsection{Laser Ionization-Injected Direct Acceleration}

Particle tracks belonging to the laser ionization-injection group, including those shown in FIG. \textcolor{blue}{\ref{fig:sample_trajectories}.b} and \textcolor{blue}{\ref{fig:sample_trajectories}.e}, are simple to identify.  As in the figure, the laser ionization-injected super-hots are created in the low density plasma a large distance from the critical surface. The critical surface is the density contour where the laser frequency is equal to the relativistic plasma frequency $\omega_{p}=\sqrt{\frac{ne^{2}}{\gamma_{L} m_{e}\epsilon_{0}}}$, where $n$ is the electron density, $\epsilon_{0}$ is the vacuum permittivity, $\gamma_{L}=\sqrt{1+\frac{a_{0}^{2}}{2}}$, (with the subscript 0 referring to the peak value of $a$).  The ionized electrons are effectively injected at the point of their creation and are accelerated towards the critical surface by the laser. Their momentum space closely follows the quadratic relationship $p_{z} = p_{x}^{2} / 2m_{e}c$ which follows from the two relations given in equation \textcolor{blue}{\ref{eq:momenta_rel}}. 

As can be seen in FIG. \textcolor{blue}{\ref{fig:sample_trajectories}.b} (as well as in the LIDA tracks shown in FIG. \textcolor{blue}{ \ref{fig:sample_trajectories}.a}), the acceleration is limited by the plasma. The dispersion relation for electromagnetic waves in plasma is $\omega^{2}=c^{2}k^{2}+\omega_{p}^{2}$ \textcolor{blue}{\cite{Kruer}} and the index of refraction is $\eta=\sqrt{1-\frac{\omega_{p}^{2}}{\omega^{2}}}$; the electromagnetic wave phase velocity $v_{p}=\frac{c}{\eta}$ is therefore superluminal. In laser ionization-injection (and LIDA), super-hot electrons are injected into the laser at relatively low density (compared to critical density, thus $\frac{\omega_{p}^{2}}{\omega^{2}}\simeq0$) where $\eta\simeq1$ and $v_{p}\simeq c$. However, as the electron is accelerated to higher density, $v_{p}$ becomes significantly superluminal and outpaces the electron whose speed is limited to $<c$. Eventually, the electron falls far enough behind that it experiences a sign reversal of the laser field causing energy loss as seen in the $z=[-18,-15]\mu m$ region (roughly) of the trajectories in FIG. \textcolor{blue}{\ref{fig:sample_trajectories}.a} and \textcolor{blue}{\ref{fig:sample_trajectories}.b}). 

of laser ionization-injected super-hot electrons originate in the lowest density region of the plasma. This is necessary for the highest energy particles since they require a relatively long acceleration length. However, there is a noticeable lack of laser ionization-injected particles at higher densities, which is unexplained.
 
\subsection{Loop-Injected Direct Acceleration}

The majority of the LIDA electrons start in the region near the critical surface (as shown by the green colored track in FIG. \textcolor{blue}{\ref{fig:sample_trajectories}.a} and \textcolor{blue}{\ref{fig:sample_trajectories}.d}), though some are swept there by the leading edge of the pulse (as shown by the red and blue colored tracks in the same figure). While ponderomotive pressure will, on average, expel electrons from the central region of the laser, it is a low frequency effect that occurs in the plasma over many laser cycles; individual tracks, like the red and blue ones, don't necessarily exhibit this average behavior. The strongest fields in the region near the critical surface are a combination of the incident laser, reflected laser, and charge separation fields. As a result, the individual particle tracks in the heating stage are complex and stochastic. Furthermore, the critical surface is complicated by two intensity effects. First, laser ionization moves the critical surface away from solid density as ions are ionized to higher stages. Second, the critical surface is dynamically shaped \textcolor{blue}{\cite{Schumacher:2011,Ping:2012}} by relativistic transparency \textcolor{blue}{\cite{Lefebvre:1995}}. The superposition of these processes creates a temporally and spatially evolving critical surface and a stochastic environment which expels electrons, some of which will become the super-hots. LIDA electrons typically leave the critical surface region with energies in the range of $10-20MeV$ for these simulation conditions; these energies are selected by the strength of the magnetic field which creates the looping.

LIDA electrons then leave the region near the critical surface and start looping in the directions indicated in FIG. \textcolor{blue}{\ref{fig:loop_cartoon}.b} and FIG. \textcolor{blue}{\ref{fig:sample_trajectories}.d} due to the azimuthal magnetic fields. These fields select the LIDA electrons from the population of electrons leaving the critical surface. The LIDA tracks in FIG. \textcolor{blue}{\ref{fig:sample_trajectories}} all have a significant transverse component to their motion $|p_{x}|>0$ and are deflected back to $x=0$. Electrons leaving the critical surface region with too much or too little energy are not injected since they are deflected either too little or too much to appropriately re-enter the most intense region of the laser field. The loop is completed when electrons transition to DLA as indicated by the acceleration to high energy in FIG. \textcolor{blue}{\ref{fig:sample_trajectories}}. 

\begin{figure}
\includegraphics[width=8cm,natwidth=800,natheight=800]{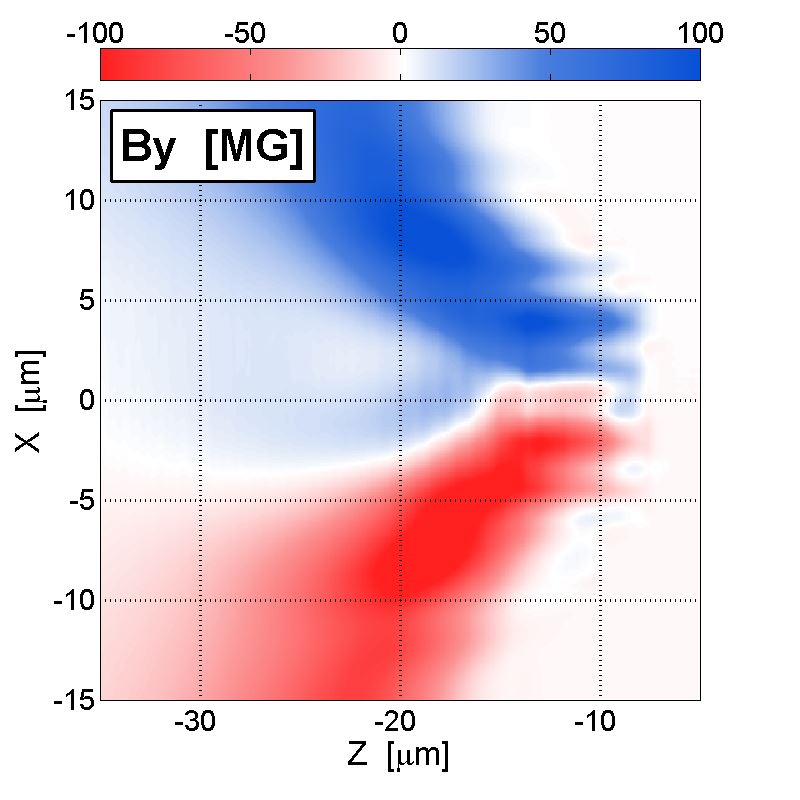}
\caption{(color online). The time ($\frac{15\lambda}{c}$) and space ($6\lambda \times 6\lambda$ moving box) averaged azimuthal magnetic field at the time when the typical super-hot is roughly halfway through its loop. The radius of curvature of a $15MeV$ electron moving in a uniform $75MG$ magnetic field is $\sim 7\mu m$ which is consistent with the observed particle tracks.}
\label{fig:By}
\end{figure}

The loop stage motion of the super-hots is controlled by the large quasi-static azimuthal magnetic field, $B_{y}$, shown in FIG. \textcolor{blue}{\ref{fig:By}} which is temporally ($\frac{15\lambda}{c}$) and spatially ($6\lambda\times6\lambda$ moving box) averaged to filter out rapidly varying fields. The quasi-static $B_{y}$ is produced by the $\vec{\nabla} T\times\vec{\nabla} n$ effect \textcolor{blue}{\cite{Stamper:1971,Mason:1998}} ($T$ is the electron temperature) which arises from the curl of the electric field caused by the plasma electron pressure gradient; the sign of these fields is opposite to that of self-generated fields due to the ponderomotive expulsion of electrons, which do not appear to play a major role in LIDA. The looping electrons typically have $U_{e}=[10,20]MeV$ during the loop stage (as shown in FIG. \textcolor{blue}{\ref{fig:sample_trajectories}.a}); for reference, the radius of curvature of a $15MeV$ electron moving in a uniform $75MG$ magnetic field is $\sim 7\mu m$ which is consistent with the observed particle tracks.

Injection into an intense laser is generally most effective for $p_{0}/m_{e}c\ll a_{0}$, where $p_{0}$ is the initial electron momentum, since large initial electron momenta can cause rapid dephasing of the electron from the laser. The need for a mechanism like LIDA arises due to the heating of the plasma in the intense field region. Looking in the general region when and where LIDA electrons are injected ($z=[-35,-25]\mu m$, $x=[-5,5]\mu m$), the average kinetic energy is $\sim 14MeV$ and only $\sim0.2\%$ of these electrons have $U_{e}<1MeV$. This explains why the background electrons, which are vastly larger in number, do not undergo efficient DLA while the looping electrons do. The plasma is so warm that efficient DLA is not possible for plasma electrons already in the high field regions because they quickly fall behind the phase of the laser due to their initial large momentum. \emph{It is apparent that the primary function of the LIDA mechanism is as an injection mechanism which delivers low energy electrons to the most intense part of the laser at the correct phase.} Most of the looping stage LIDA electrons are \emph{initially} too energetic to undergo efficient DLA. However, as the electrons loop into the high field region of the laser, they move against the ponderomotive potential of the laser which dissipates a significant fraction of their energy. In general, electrons with too little energy, including those not participating in the loop but in the same region, are turned away by the ponderomotive forces of the laser, and those with too much energy simply overshoot the acceleration region. The looping electrons that will become the super-hots lose the majority of their energy just prior to undergoing DLA. While the individual particle tracks are complicated by charge separation fields and reflected laser light, our results confirm that this history is a universal feature for the overwhelming majority of electrons that become super-hots. 

\begin{figure}
\includegraphics[width=8cm,natwidth=3978,natheight=2068]{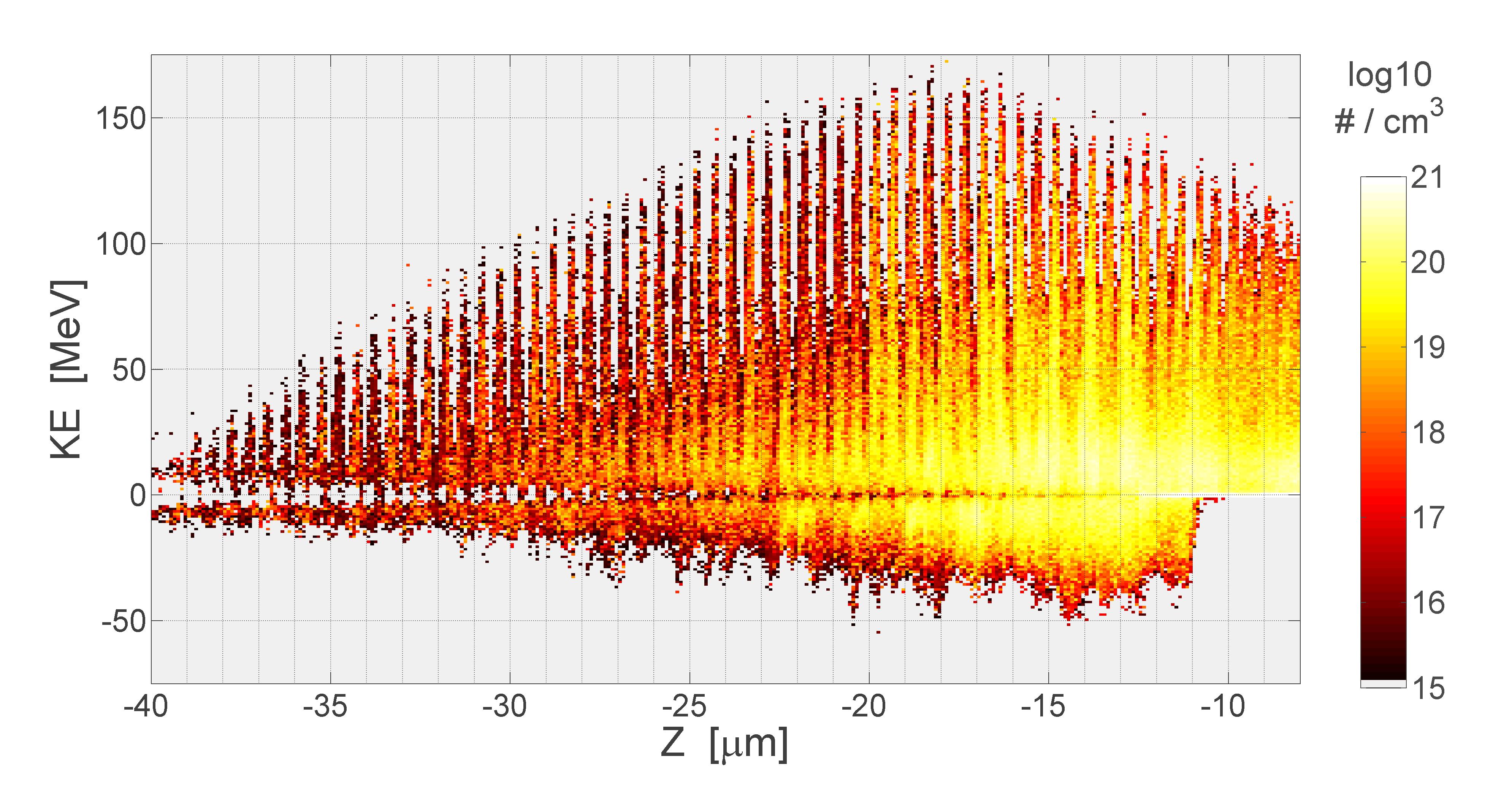}
\caption{(color online). Kinetic energy (carrying the sign of $p_{z}$) vs. longitudinal position for electrons in the approximate region where DLA starts ($x=[-5,5]\mu m$) at the time when nearly half of the super-hots have started accelerating. The regular $2\omega$ bunching indicates that the electrons are undergoing DLA.}
\label{fig:z_KE}
\end{figure}

There are several features of DLA in the super-hot particle tracks. FIG. \textcolor{blue}{\ref{fig:z_KE}} shows kinetic energy (carrying the sign of $p_{z}$) vs. longitudinal position for electrons in the range $z=[-40,-8]\mu  m$ where each position in $z$ is summed over $x=[-5,5]\mu m$. There is a pronounced $2\omega$ bunching in the forward moving hot electrons; this bunching is a well-known signature of $J\times B$ forcing and arises due to the electric and magnetic field of a laser being in phase. Furthermore, after injection, the accelerating electrons follow very closely (not shown here) the signature parabolic momentum relation, $p_{z}=p_{x}^{2}/2m_{e}c$ referred to earlier. Finally, while the electron density shows large plasma waves/charge disturbances, we don't observe any regular charge arrangements that could produce the super-hot electron energies as in the wake-field acceleration mechanism.

The spatial profile of the laser pulse also plays an important role in LIDA. On average, electrons will tend to move against the gradient of the intensity of the laser; electrons near the center of the laser are pushed to weaker field regions. However, as in the plane wave discussion, the super-hot electrons have an extended acceleration length because they remain at approximately constant phase relative to the laser. The largest energies are achieved by electrons that move towards the center of the pulse where the laser field is strongest. This is evident in the tracks shown in FIG. \textcolor{blue}{\ref{fig:sample_trajectories}.d}: the green track loops with $x>0$ and is accelerated with $p_{x}<0$, the red and blue tracks loop with $x<0$ and are accelerated with $p_{x}>0$.

\section{Discussion}

The modeling in this paper represents a small part of the total explorable parameter space. Accordingly, the results presented above are subject to change under different conditions. Here we speculate on varying a number of these parameters; however, making generalizations based on the isolation of individual parameters is difficult as the parameters are intrinsically coupled.

\subsection{Peak Intensity}

The large electron energies observed in our modeling require a large acceleration length (compared to $\lambda$). The $\vec{v}\times\vec{B}$ term in the Lorentz force law gives rise to electron motion along the laser propagation direction; as the strength of the field becomes large, the forward motion can become significant and even dominant. The transition to forward dominated motion occurs when the ratio $p_{z}/|p_{x}| = |a-a_{i}|/2$ is greater than $1$ which can occur for $a_{0}>1$. However, a large acceleration length can only occur when the longitudinal momentum becomes much larger than the transverse ($p_{z}/|p_{x}| \gg 1$).  LIDA and laser ionization-injection enable the large acceleration distances by injecting electrons in the strongest region of the laser at a large distance from the critical surface. So for moderately relativistic laser intensities, LIDA is not expected to contribute as significantly as here since the acceleration lengths are of order $\lambda$. LIDA does however play a similarly dominant role over the range of peak intensities $I=10^{20}-10^{21} \frac{W}{cm^{2}}$ ($a_{0} \simeq 7 - 27$, for $\lambda = 1\mu m$) based on simulations with otherwise identical parameters not presented herein.

\subsection{Pulse Length}

Modern short-pulse laser facilities are now able to produce high intensity laser pulses with a wide range of pulse lengths. Lasers employing titanium-sapphire amplifiers are able to produce pulses shorter than $30fs$ whereas glass amplification systems typically operate in the $500-1000fs$ range. More recently developed hybrid systems are able to produce pulses similar in length to the one used in the above modeling.  

In the case of the $\sim30fs$ pulses, LIDA should not play much of a role due to time-of-flight restrictions on the electron. An electron simply does not have the time to be ejected from the critical surface region, loop backwards and re-enter the laser pulse in time to be injected into the peak intensity region of the laser. Furthermore, the magnetic fields that control the loop stage will likely not be able to grow fast enough to produce the looping paths. Absent a new, unknown injection mechanism, the energy spectrum should be cooler since a larger proportion of the hot electrons will have $a_{i}\simeq 0$. 

For $500-1000fs$ length pulses and longer, there is no time-of-flight issue and it is possible that the role of LIDA will be maintained. This may be consistent with the findings of Kemp, et al. \textcolor{blue}{\cite{Kemp:PRL2012}}. Their modeling was slightly different from ours in that they started with a sharp interface and a $>1ps$ pulse. However, they observed a hotter energy spectrum as plasma expanded away from the sharp interface; they attribute the enhancement to direct laser acceleration of the expanding plasma electrons.

\subsection{Focusing}

The LIDA mechanism relies on the growth of $\vec{\nabla} T \times \vec{\nabla} n$ magnetic fields. By itself, the focusing determines the transverse intensity gradient and therefore the transverse thermal gradient of the electrons. Therefore, a tighter focus should produce faster growing magnetic fields with a smaller extent and softer focusing should produce slower growing magnetic fields but with a larger extent. The electrons that participate in LIDA are selected from the total population of electrons that exit the critical surface by the azimuthal magnetic field. Weak focusing will therefore select lower energy electrons for LIDA and tight focusing will select higher energy electrons.

If the looping energy is significantly reduced due to a slowly growing magnetic field, the electrons may not be able to overcome the ponderomotive potential and be scattered out of the laser focus with low energy. However, in the modeling described above, the energy distribution (integrated over all angles; not shown) of electrons in the critical surface region when the average super-hot starts to loop is well-fit by an exponential function with a "slope temperature" of $\sim 5.6 MeV$. Therefore, a lower looping energy would also substantially increase the number of electrons that could participate in LIDA.

Significantly increasing the looping energy (and decreasing the spatial extent of the magnetic field) with a tighter focus will almost certainly reduce the number of particles that can loop and eventually become super-hot via LIDA. Furthermore, increasing the energy could cause the electrons to overshoot the laser and never get injected.

Practically speaking, changing the focusing most directly alters the laser intensity. The transverse profile of the intensity of a Gaussian laser at focus scales with $e^{-r^{2}/2w_{0}^{2}}$ where $r$ is the radial distance from the focus; therefore, the peak intensity (and equivalently $a_{0}^{2}$) scales as $1/w_{0}^{2}$. As a result, both the energy an electron can gain from DLA as well as the magnetic field growth rate are a function of the focusing and it is difficult to determine how these competing effects balance without further modeling.

\subsection{Pre-plasma Scale Length}

The growth of the $\vec{\nabla} T \times \vec{\nabla} n$ magnetic fields also depends on the pre-plasma scale density gradient. In our modeling, we have used a 1D exponential decay with a cut-off for the plasma density; these magnetic fields will grow faster for shorter scale-length plasma. For this kind of pre-plasma density distribution, the scale length also defines the spatial extent of the plasma. We performed several calculations (not shown) to investigate the dependence of LIDA on the scale length and found that the role of LIDA is similar to the above results when $L=5\mu m$ but is significantly diminished when $L=1\mu m$.

As the scale length is increased from $L=3\mu m$, LIDA is initially expected to produce even higher energy electrons but will eventually diminish due to weaker magnetic fields and a larger region where $v_{p}$ is significantly different than $c$. However, it is reported by \textcolor{blue}{\cite{Pukhov:1999}} that longer scale-lengths (though with a less intense laser) produce hotter distributions. In that case, the hotter distributions were reported to be produced by the betatron resonance mechanism. In the betatron resonance mechanism, electrons undergoing DLA are also forced by self-generated magnetic fields and electrostatic fields due to the ion channel formed by the ponderomotive expulsion of electrons. Both of these fields will radially pinch accelerating electrons; as explained in \textcolor{blue}{\cite{Pukhov:1999}}, "when the frequency of the transverse electron oscillations in the self-generated static electric and magnetic fields (betatron oscillations) coincides with the laser frequency as witnessed by the relativistic electron, a resonance occurs". More recently, Arefiev \textcolor{blue}{\cite{Arefiev:2012}} has shown that the frequency of electron oscillations across the channel can be strongly modulated by the laser field leading to parametric energy gain. This is distinct from the betatron resonance because it can occur out of the plane of polarization of the laser. However, LIDA and these acceleration mechanisms are not mutually exclusive since LIDA is an injection mechanism and the others are acceleration mechanisms.

For smaller scale lengths than considered in our modeling, the magnetic fields are expected to grow larger, but with smaller extent. As with the intensity gradients, this will have the effect of reducing the number of electrons which can participate in the process. Furthermore, the plasma will not support long acceleration lengths since the magnetic field does not extend into the vacuum.

\section{Conclusions}

DLA is found to be the dominant acceleration mechanism for the generation of super-hot electrons when an ultra-intense laser interacts with a solid with moderate scale-length pre-formed plasma. We have studied conditions similar to recent short pulse laser experiments using flat targets and we find that the majority of the super-hot electrons require a loop-like injection mechanism we call LIDA. This work could have implications for ion acceleration, especially hadron cancer therapy as well as bright x-ray production which are both enhanced for a hotter electron energy spectrum. Our findings should lead to improved hot electron coupling in a number of ways. For instance, a pre-plasma that enhances acceleration length and/or minimizes phase velocity mismatch may be possible with the development of advanced targets specifically designed to enhance the delivery of looping electrons.

\section{Acknowledgments}

We thank J.T. Morrison for discussions. This work was supported in part by an allocation of computing time from the Ohio Supercomputer Center. This work was performed under the auspices of US Department of Energy under contracts DE-FC02-04ER54789, DE-FG02-05ER54834 and DE-NA0001976.

\bibliography{bibfile_POP}

\end{document}